\documentclass[sigconf]{acmart}

 \author{Anne Müller}
 \orcid{0000-0002-4468-2651}
 \affiliation{
  \institution{CISPA Helmholtz Center for Information Security}
  \city{Saarbrücken}
  \country{Germany}
}
\affiliation{
  \institution{Graduate School of Computer Science, Saarland University}
  \country{Germany}
}
 \email{anne.mueller@cispa.de}

 \author{Mohd Kashif}
 \affiliation{
  \institution{CISPA Helmholtz Center for Information Security}
  \city{Saarbrücken}
  \country{Germany}
}
\affiliation{
  \institution{Saarland University}
  \country{Germany}
}
 \email{md.kashif8858@gmail.com}

 \author{Nico Döttling}
 \authornote{Funded by the European Union (ERC, LACONIC, 101041207). 
Views and opinions expressed are those of the author(s) only and 
do not necessarily reflect those of the European Union or the ERC. 
Neither the European Union nor the granting authority can be held responsible.}
 \affiliation{
  \institution{CISPA Helmholtz Center for Information Security}
  \city{Saarbrücken}
  \country{Germany}
} \email{doettling@cispa.de}

\usepackage{enumitem}
\usepackage{algorithm}
\usepackage{algpseudocode}
\usepackage{booktabs}   % for \toprule, \midrule, \bottomrule
\usepackage{array} 
\usepackage{cleveref}
\usepackage{tikz}
\usetikzlibrary{positioning}
\newcommand{\Keygen}{\mathsf{KeyGen}}
\newcommand{\Enc}{\mathsf{Enc}}
\newcommand{\Dec}{\mathsf{Dec}}
\newcommand{\Eval}{\mathsf{Eval}}
\newcommand{\pk}{\mathsf{pk}}
\newcommand{\sk}{\mathsf{sk}}
\newcommand{\ct}{\mathsf{ct}}
\newcommand{\msg}{\mathsf{m}}
\newcommand{\servantoine}{\texttt{AMD EPYC}}

\begin{document}

\acmYear{2025}\copyrightyear{2025}
\setcopyright{acmlicensed}
\acmConference[WAHC '25]{Proceedings of the 2025 Workshop on Applied
Homomorphic Computing}{October 13--17, 2025}{Taipei, Taiwan}
\acmBooktitle{Proceedings of the 2025 Workshop on Applied Homomorphic Computing (WAHC '25), October 13--17, 2025, Taipei, Taiwan}
\acmDOI{10.1145/3733811.3767311} %% CHECK
\acmISBN{979-8-4007-1900-4/25/10} 

\title{A Haskell to FHE Transpiler With Circuit Parallelisation} 

\begin{abstract}
Fully Homomorphic Encryption (FHE) enables the evaluation of programs directly on encrypted data. However, because only basic operations can be performed on ciphertexts, programs must be expressed as boolean or arithmetic circuits. This low-level representation makes implementing applications for FHE significantly more cumbersome than writing code in a high-level language.

To reduce this burden, several transpilers have been developed that translate high-level code into circuit representations. In this work, we extend the range of high-level languages that can target FHE by introducing a transpiler for Haskell, which converts Haskell programs into Boolean circuits suitable for homomorphic evaluation.

Our second contribution is the automatic parallelization of these generated circuits. We implement an evaluator that executes gates in parallel by parallelizing each layer of the circuit. We demonstrate the effectiveness of our approach on two key applications: Private Information Retrieval (PIR) and the AES encryption standard. 
Prior work has parallelized AES encryption manually. 
%While previous work has manually parallelized AES encryption,
%We demonstrate that the automated method achieves comparable or better performance for AES evaluations using Boolean circuits. 
We demonstrate that the automated method outperforms some but not all manual parallelizations of AES evaluations under FHE. We achieve an evaluation time of 28 seconds for a parallel execution with 16 threads and an evaluation time of 8 seconds for a parallel execution with 100 threads.   
\end{abstract}

% TODO: replace this section with code generated by the tool at https://dl.acm.org/ccs.cfm


 % TODO: replace with your keywords

\maketitle

\section{Introduction}
Fully Homomorphic Encryption (FHE) is a foundational tool for secure cloud computing and privacy-preserving machine learning. Since Gentry’s groundbreaking work in 2009 \cite{gentry09}, the field has developed rapidly.  However, the broad adoption of FHE still faces several challenges. Most notably, the computation on encrypted data is very inefficient compared to the computation on plain data. Moreover, only basic operations are possible on ciphertexts. Therefore, the program has to be specified as a Boolean or arithmetic circuit. 

To mitigate the aforementioned issues many improvements in the area of homomorphic encryption have been introduced. New schemes have been developed that achieve better efficiency. Additionally, many compilers have been developed that transform code from a high-level language such as C++ or Python into a low-level circuit representation~\cite{SOK}. 

However, many of these compilers support only a restricted subset of their respective source languages. In contrast, our approach starts with programs written in Clash, a functional hardware description language that supports most features of Haskell. Naturally, FHE-specific constraints still apply, such as no data-dependent recursion. After the compilation step we evaluate the circuit using the TFHE library~\cite{TFHElib}. 

The issue of low efficiency can be addressed by parallelizing FHE programs. But, such parallelism cannot typically be implemented in the source language: to the best of our knowledge, no current compiler supports automatic translation of libraries containing parallel execution instructions. As a result, developers have manually identified parallelizable code regions, compiled them separately, and reassembled them for parallel circuit evaluation—a cumbersome, error-prone process that must be repeated for each new application.

Contrary to this approach, we investigate the performance of simple layer-based parallelization. A circuit naturally consists of layers such that each gate in a layer can be evaluated independently of the gates in the same layer. 

We test our compiler on an AES implementation. AES is a common benchmark for FHE implementations and a popular target for manual parallelization efforts. Trama et al.~\cite{TCBS23} report an evaluation of AES using 16 threads in 36 seconds on an AMD-server or 28 seconds on an i7-server. Our evaluation similarly achieves an AES evaluation in 28 seconds using 16 threads. The choice of 16 threads for parallelizing AES is no coincidence. The state throughout the AES execution consists of 16 bytes and in each round some of the operations can be performed simultaneously on each byte of the state. Therefore, for a manual parallelization of AES the choice of 16 threads is natural. Our parallelization algorithm is oblivious to any structure other than the number of gates at each depth level of the circuit. Therefore, the parallel execution can be extended to any number of threads; as such we achieve an evaluation time of 8 seconds using 100 threads. 

In a recent paper \cite{SNARDN25} the 'Hyppogryph' implementation of AES achieves a single-threaded execution in 32 seconds and an execution with 32 threads in 1.1 seconds. The implementation is highly specialized and combines multiple methods for different sections of the AES cipher. Therefore, the manual parallelization in this approach seems necessary.

\textbf{Contributions}
\begin{itemize}
    \item  We present the first compiler that translates Haskell (via Clash) into circuits suitable for fully homomorphic encryption, making FHE accessible to a broader audience, particularly those in the functional programming community. 
    \item As our second contribution, we show that the natural circuit parallelization technique achieves good results on important benchmarks such as AES and should not be neglected. We further evaluate our compiler on a second benchmark where we implement private information retrieval (PIR). 
\end{itemize}

The code, including all benchmarks, is publicly available\footnote{https://github.com/Anne-Me/Haskell2FHE}. 

\section{Background}

\subsection{Fully Homomorphic Encryption}
A fully homomorphic encryption scheme consists of the following polynomial time algorithms: 
\begin{itemize}[align=left,leftmargin=4.2em]
    \item[$\mathsf{KeyGen}(1^{\lambda}) \rightarrow (\pk,\sk)$] $\Keygen$ is a probabilitstic algorithm that on input the security parameter $\lambda$ produces a public key, secret key pair $(\pk,\sk)$.
    \item[$\mathsf{Enc}(\pk,\msg) \rightarrow \ct$] $\Enc$ is a probabilisitc algorithm that on input the public key $\pk$ and a message $\msg$ produces a ciphertext $\ct$. 
    \item[$\mathsf{Eval}(\pk, \mathcal{C}, \ct)\rightarrow \ct'$] $\Eval$ is a probabilistic algorithm that on input the public key $\pk$ and a circuit $\mathcal{C}$ and a ciphertext $\ct$ produces a ciphertext $\ct'$.
    \item[$\mathsf{Dec}(\sk,\ct) \rightarrow \msg / \bot$] $\Dec$ is a deterministic algorithm that given a secret key $\sk$ and a ciphertext $\ct$ outputs a message $\msg$ or $\bot$.
\end{itemize}

The FHE scheme also has to fulfill security and efficiency properties. Informally, we require that given a ciphertext, one cannot learn anything about the encrypted message. We also require that the size of the ciphertext that is produced by Eval does not depend on the size of the circuit $\mathcal{C}$. Of course, we also require correctness, which means that given an encryption of a message $\msg$, after applying $\mathsf{Eval}(\pk, \mathcal{C}, \ct)\rightarrow \ct'$ the resulting ciphertext $\ct'$ contains an encryption of $\mathcal{C}(\msg)$. Note that the circuit $\mathcal{C}$ is not hidden. 

The circuit $\mathcal{C}$ that can be evaluated by the $\Eval$ algorithm depends on the particular FHE scheme. For example, GSW~\cite{C:GSW13}, FHEW~\cite{EC:DucMic15} and TFHE~\cite{TFHE} allow the execution of boolean circuits while BGV~\cite{BGV} and BFV~\cite{brakerski12,FV} allow the evaluation of arithmetic circuits with modular arithmetic. The CKKS~\cite{AC:CKKS17} scheme allows to evaluate arithmetic circuits on real numbers. In this work we make use of the TFHE library~\cite{TFHElib} which works on boolean circuits. 

A levelled homomorphic encryption scheme is a homomorphic encryption scheme that can only evaluate an a priori bounded number of gates. The number of gates has to be known at key-generation time. This is in contrast to a fully homomorphic encryption scheme which can evaluate an unbounded number of gates on a ciphertext. This is achieved through an expensive bootstrapping operation. With each homomorphic operation, noise accumulates in the ciphertext, which would eventually lead to decryption errors. To avoid this the error has to be reduced periodically by applying the bootstrapping algorithm which is a time-consuming operation. In the FHEW scheme the cost of the bostrapping operation was amortized by combining the boostrapping with a NAND gate. This operation was further improved in its efficiency in the TFHE scheme. A ciphertext of these schemes can also encrypt multiple bits. Then, the functional boostrapping technique allows to evaluate a small boolean function with a few input and output bits, typically 3-8, within a boostrapping operation. 

%textcolor{red}{LUT tables stuff} multi-value functional %boostrapping\cite{SMV19}\cite{kluczniak2021fdfb}

\subsection{Circuit Representation of FHE Programs}
A circuit can be represented as a directed acyclic graph $G = (V,E)$ where $V$ is the set of vertices in the graph and $E$ is the set of edges. An edge indicates a wire in the circuit. A vertex $v$ is annotated with a gate, for example in the case of boolean circuits the gate can be \texttt{AND,OR,NAND,XOR}. We call $v_1$ a \textit{parent} of $v_2$ if there is an edge from $v_1$ to $v_2$. We call $v_2$ a \textit{child} of $v_1$ if there is an edge going from $v_1$ to $v_2$. 

The depth of a node is defined as the maximum length path one can take from an 'input' node, i.e. a node that has no incoming edges to the node. The depth of the DAG is defined as the maximum depth of all nodes. 

%A subcircuit is defined by a subset of the vertices of the graph and contains all edges between the vertices within the subcircuit. Subcircuits are called independent if there are no edges going from one subcircuit to the other. Independent subcircuits can be evaluated in parallel since none of the evaluation steps in one subcircuit depends on the other subcircuit. 

\subsection{Related Work}

\subsubsection{Automatic Parallelisation}
Automatic parallelization is, in general, a very difficult task since most programs rely on data-dependent branching and loops. Then, possible opportunities for parallelisation have to be identified dynamically at runtime~\cite{SHZH12}. In FHE programs however no such data-dependent operations are allowed, making it a prime target for automated parallelisation. The question of optimally parallelising FHE computations can be defined as the task of splitting a circuit into equally sized, independent subcircuits. In general the task of dividing a circuit into k subcircuits of equal size such that the number of edges between subgraphs is minimised is NP-hard~\cite{AndreevHarald04}. We employ a not necessarily optimal but natural parallelisation technique for circuits by parallelising each layer of the circuit individually.

\subsubsection{FHE Compilers} With the progress in FHE schemes also comes progress in the FHE compiler development literature such that the new and improved features of FHE schemes can be leveraged. Among the first FHE compilers were Cingulata \cite{cingulata}, Alchemy \cite{alchemy}, Marble \cite{marble},  E3 \cite{E3} and Ramparts \cite{ramparts} which take high-level code and transform it into arithmetic or boolean circuits to be executed on BGV or GSW ciphertexts. An important optimization technique that has received much attention is packing or SIMD batching which allows to pack multiple inputs into one ciphertext. Some compilers that implement SIMD batching require the user to provide the input in vectorized format \cite{ramparts,eva}. To reduce the effort and expertise required by the user progress was made to automatically translate SIMD-friendly programs by the compilers HECO, Fhelipe, HEIR and Porcupine ~\cite{heco, fhelipe, porcupine,HEIR}. 
%The choice of compiling to arithmetic or boolean circuits and subsequently the choice of FHE scheme to evaluate the circuit can depend on the application. For some applications, arithmetic circuits have better performance, while for other applications boolean circuits have superior performance. Therefore, the HEIR \cite{HEIR} library implements a transciphering operation that translates a ciphertext from one scheme into a ciphertext from another scheme. Then part of the program can be compiled to a boolean circuit and another part can be compiled into an arithmetic circuit. 
% elasm \cite{elasm}, 
EVA \cite{eva}, Ant-Ace \cite{antace} and nGraph-HE \cite{nGraph-HE} are compiler frameworks that are developed with a focus on machine-learning applications such as encrypted interference. They evaluate the finalised circuits under CKKS since its inherent support for real numbers makes it practical for machine-learning tasks. 
%CHET \cite{chet} also is 

The functional bootstrapping approach requires the creation of circuits consisting of LUTs (look-up tables) instead of boolean or arithmetic gates. LUTs have long been a fundamental concept in hardware-accelerated computation, such as in FPGAs (Field-Programmable Gate Arrays). Therefore, many compilers can easily be extended to compile to LUT tables instead of boolean circuits by leveraging hardware synthesis tools such as Yosys. The HEIR \cite{HEIR} compiler and the compiler by Mono et al.~\cite{MKG24} support LUT tables via Yosys compilation. 
%Bon et el. \cite{BPR24} employ a different approach to transform a boolean circuit into a LUT-based circuit. On small specialized circuits for cryptographic applications, they exhaustively search possible transformations from boolean functions to LUT tables or they apply a heuristic for larger circuits. 

%Gates with larger fan-in \cite{KYTS21,MSCD24} 
%The input and output size of the LUT table is another parameter to consider when transforming a boolean circuit into a LUT table based circuit. 

In the domain of arithmetic circuit optimisation Oraqle \cite{oraqle} performs depth-aware restructuring of the circuit by trying to balance the multiplicative depth which is responsible for noise growth with the multiplication count which increases the runtime. 
\section{The Transpiler}
\begin{figure}
    \centering
    \begin{tikzpicture}[node distance=1.5cm, thick]
    \node[draw, rectangle, minimum width=2.2cm, minimum height=1cm] (A) {Clash};
    \node[draw, rectangle, minimum width=2.2cm, minimum height=1cm, right=2cm of A] (B) {Yosys};
    \node[draw, rectangle, minimum width=2.2cm, minimum height=1cm, below=of B] (C) {Paralleliser};
    \node[draw, rectangle, minimum width=2.2cm, minimum height=1cm, left=2cm of C, align=center] (D) {Parallel\\ Evaluation};

    \node[coordinate, left=of A] (In) {};
    \draw[->] (In) -- node[above] {Haskell} (A);
    \draw[->] (A) -- node[above] {HDL} node[below] {(Verilog)} (B);
    \draw[->] (B) -- node[right, align=center] {Boolean\\Circuit}(C);
    \draw[->] (C) -- node[above, align=center] {Boolean} node[below, align=center] {Circuit}(D);
\end{tikzpicture}
    \caption{Our synthesis toolchain}
    \label{fig:toolchain}
\end{figure}
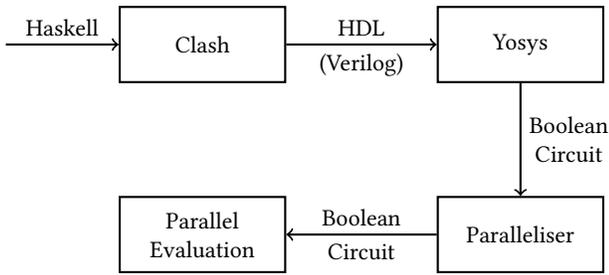

The toolchain of our transpiler consists of four main stages which are depicted in \cref{fig:toolchain}. First, the code written in the high-level language Haskell is compiled into the hardware description language (HDL) Verilog using the Clash\footnote{https://clash-lang.org/} compiler. Then, the verilog code undergoes optimisations and is translated into a boolean circuit using Yosys\footnote{https://github.com/yosyshq/yosys}. This boolean circuit enters the parallelization stage of our compiler where the circuit is prepared for the parallel execution. This stage is described in more detail in \ref{sec:level-parallel}. Finally, the circuit is evaluated in a multi-threaded manner by the Evaluation procedure which uses the TFHE library.

The toolchain is similar to other compilers such as HEIR \cite{HEIR} which also leverages hardware language synthesis and optimization tools. HEIR  focuses on creating the program in a  Multi-Level Intermediate Representation (MIRL) which is then optimized and compiled into Boolean circuits. We opted to use the compiler Clash since it supports native Haskell code. A programmer who is familiar with Haskell can immediately start implementing and does not need familiarize themselves with a new language. Of course, some limitations that are inherent to the representation of programs in circuits apply. Clash does not support data-dependent recursion or recursive data types, therefore only a subset of the Haskell language is supported. 

Yosys performs many optimisations such as constant-folding, dead-code elimination and depth reduction. In particular, depth reduction is useful for the following step in our workflow since it increases the number of gates that can be executed in parallel. 
%\textcolor{red}{exmaple with XOR tree?}
%\textcolor{red}{example with haskell code?}

\subsection{Parallelization}
\label{sec:level-parallel}
To prepare the circuit for parallel execution each gate is assigned a depth. We note that the term depth might be used differently in different works. In the context of FHE circuits depth often refers to the multiplicative depth of the circuit, i.e. the maximum number of sequential multiplications. This is usually a relevant measure due to the noise growth which is stronger in multiplication operations than in addition operations. 
In our application we need to identify the layer in which the gate sits so that we know at which point it can be executed. The depth of a node is the maximum of all its parents' depths + 1. This ensures that a gate is only evaluated after its parents have been evaluated. To assign a depth to the gate we first compute a topological ordering of all gates. In a topological sorting a node $v_1$ appears before a node $v_2$ if there is an edge from $v_1$ to $v_2$. The topological sorting can be computed in time O(|V|+|E|) using depth-first search. Start at an input node and continue through the graph according to the depth-first search. Upon reaching a terminal node, i.e. a node with no outgoing edges or a node where all child nodes are already collected, append the node to the  list of sorted gates and continue. After all nodes are collected, assign depth 0 to all input nodes. Then traverse the list of topologically sorted nodes and for every node $v_1$ if $v_1.depth = i$ then for all children $v_2$ of $v_1$ set $v_2.depth =$ max$\{ i+1, v_2.depth\}$.

During evaluation we are given a number of threads $k$ and for each layer all gates are split equally between the available threads. 

\section{Applications}
In this section, we evaluate our approach on two important FHE applications:  AES encryption and Private Information Retrieval (PIR). We evaluat the applications on a AMD EPYC 9374F 32-Core Processor with 128 available vCPUs running a Ubuntu 24.04.2 LTS server. We instantiated the TFHE library with a security level of 128 bits.

\subsection{AES}

AES is a symmetric encryption algorithm which was standardized by NIST in 2001. It operates on 128-bit blocks of plaintext and produces 128-bit ciphertexts. The algorithm supports key sizes of 128, 192, or 256 bits. In this benchmark, we focus exclusively on \textbf{AES-128}, which uses a 128-bit key.
The algorithm consists of two main phases: 1. Key Expansion and 2. Encryption. 
In the key expansion step the 128-bit key is expanded into 11 round keys each of size 128 bits. The encryption phase of AES begins by XORing the plaintext to the first round key. For the following steps it is useful to imagine the 128 bit state to be split into 16 byte-sized words which are arranged in a 4-by-4 matrix. Then the algorithm proceeds in 9 rounds where each round consists of the following steps
\begin{enumerate}
    \item SubBytes: A non-linear substitution where each byte is replaced according to a fixed S-box.

    \item ShiftRows: Rotate the last three rows of the state by a fixed number of bytes.

    \item MixColumns: A linear mixing operation combining bytes within one column.

    \item AddRoundKey: XOR the current state with the corresponding round key.
\end{enumerate}
One final round is added where only SubBytes, ShiftRows and AddRoundKey are applied. 

While traditional AES implementations often use a lookup table for the S-box, this approach is suboptimal in FHE settings. Instead we implement the S-box using a Boyar-Peralta circuit~\cite{bpsbox}. This circuit uses only 128 gates to implement the S-box.

Due to the large ciphertext size in a FHE scheme it is preferable to send the inputs of a computation encrypted under the AES block cipher. The AES key is encrypted under the FHE scheme such that the inputs can be homomorphically decrypted on the server side. Although more FHE-friendly ciphers have been proposed for practical use, AES remains an important benchmark.

Following prior work~\cite{stracovsky22,TCBS23}, we benchmark only the encryption phase of AES, as the key expansion can be performed in advance by the client. Our transpiler produces a circuit with 27987 boolean gates. In \cref{tab:AES-times} we report the execution times of previous works and our work. 

The work of Mella and Susella \cite{MS13} implements AES and other cryptographic algorithms such as SHA-256, Salsa20 and KECCAK. Their implementation targets the BGV scheme. The implementation of \cite{C:CSN12} also targets the BGV variant and they report execution times of AES under FHE in 22 minutes. Additionally, they achieve an execution of AES in 4 minutes in a levelled homomorphic encryptino scheme, therefore, the ciphertext cannot be used in further computations after the AES execution.  
In the domain of levelled homomorphic encryption \cite{fregata} and \cite{thunderbird} achieve strong improvements which we summarize in \cref{tab:AES-levelled}. 
The work of Trama et al.~\cite{TCBS23} and Bon et al.~\cite{BPR24} target LUT based TFHE implementations. Bon et al.~\cite{BPR24} implement a general framework to decompose a boolean circuit into boolean functions that can be implemented as a LUT table. They apply a p-encoding approach to encode a bit with redundancy. The gadgets for boolean functions can be applied which reduces the number of functional bootstrappings necessary for the evaluation of the function. To find the gadgets for a boolean function and to find partial boolean functions in a larger circuit they develop heuristic algorithms. They also discuss parallelization opportunities for their scheme but they do not report execution times. They mention the well-known 16-thread parallelization technique and also suggest the layer-based parallelization technique to be applied locally within one S-Box execution.

Trama et al.~\cite{TCBS23} focus on transforming the steps of each AES round into 8-by-8-bit or 4-by-4-bit LUT tables to match the structure of the AES byte structure. Then they manually parallelize the SubBytes, MixColumns and AddRoundKey functions in each round of the AES encryption step. We demonstrate that we can automatically parallelize the circuit based on the layer-parallelization approach and achieve similar execution times. Both the approach \cite{TCBS23} and our automatic parlallelization achieve an execution time of 28 seconds for 16 gates. Furthermore the layer-paralellization approach yields good results for any number of available threads while the manual parallelization approach does not allow a variable number of threads. We have tested the AES execution up to 100 threads and achieve an execution in 8.447 seconds.

In a recent work \cite{SNARDN25} achieve a strong improvement by combining the LUT-based techniques of Trama et al.~\cite{TCBS23} and the p-encoding technique of Bon et al.~\cite{BPR24} in an implementation they name Hippogryph. This approach is highly specialized to the AES circuit and requires encoding switching between the roundkeys and the subBytes steps of each round in the AES implementation. Their results are included in \cref{tab:AES-times}.

\begin{table}[ht]
\centering
\begin{tabular}{l|c|c}

& \textbf{Threads} & \textbf{Time} \\
\midrule
Gentry et al.\ \cite{C:CSN12}        & 1   & 18min \\
Mella et al.\cite{MS13}      & 1     & 22 min\\
Stracovsky et al.\ \cite{stracovsky22}  & 16  & 4.2min \\
Bon et al. \cite{BPR24} & 1 & 135s\\
\midrule
Trama et al. \cite{TCBS23}& & \\
\midrule
%\texttt{i7-laptop}  & 1  & 270s (4.5min)  \\
%\texttt{i7-laptop}    & 6           & 54.31s  \\
\texttt{AMD-server}     & 1  & 342s (5.7min) \\
\texttt{AMD-server}   & 16            & 36.39s  \\
\texttt{i7-server} 
%\footnote{The time reported for the i7-server is an estimation based on the AMD-server and the i7-laptop since the authors did not have an i7 server available.} 
& 16          & 28.73s  \\
\midrule
Hyppogryph \cite{SNARDN25}& 1 & 32s\\
Hyppogryph \cite{SNARDN25} & 32 & 1.1s\\
\midrule
Our work & & \\
\midrule
\servantoine & 1 & 311.301s (5.18min) \\
%\servantoine & 6 & 69.594s\\
\servantoine & 16 & 28.359s\\
\servantoine & 20 & 23.596s\\
\servantoine & 50 & 11.046s\\
\servantoine & 100& 8.447s\\
%laptop? & & \\
\bottomrule
\end{tabular}
\caption{Our execution times of AES-128 and those of previous works.}
\label{tab:AES-times}
\end{table}

\begin{table}[ht]
\centering
\begin{tabular}{l|c|c}

& \textbf{Threads} & \textbf{Time} \\
\midrule
Gentry et. al. \cite{C:CSN12}  & 1& 4min\\
Fregata  \cite{fregata}  & 1     & 86s\\
Fregata   \cite{fregata}     & 16   & 9s \\
Thunderbird \cite{thunderbird} & 1 & 46s \\
\bottomrule
\end{tabular}
\caption{Execution times for AES encryption in levelled homomorphic encryption schemes.}
\label{tab:AES-levelled}
\end{table}

\paragraph{How much can AES be parallelized?} Due to the narrow circuit structure of AES a much bigger speedup due to parallelization only cannot be expected. The AES circuit produced by our scheme has a depth of 204. In all but 1 layer there are less than 200 gates and more than 50\% of the layers have fewer than 150 gates. Since we can have at most as many parallel threads as we have gates in a layer, the maximum speed-up could be achieved for 200 threads. Of course, creating threads and waiting for threads to join can create additional overhead. Therefore, more than 100 gates will likely not yield a much bigger improvement and different compilation techniques have to be applied. 
%More sophisticated advancements as in \cite{moreAES} are necessary to improve the timings

\subsection{Private Information Retrieval}
Imagine a hospital has a database with patient records that a research team wants to access. The hospital needs to keep its records on the server encrypted to comply with regulations and the research team does not want to reveal which illness they are researching. This scenario can be solved using Private Information Retrieval (PIR). PIR is the task of obliviously selecting a database entry based on the client's request without learning anything about the client's data. In the real world, Microsoft Edge’s Password Monitor \cite{psswordmonitor} uses FHE to check whether a user’s saved passwords appear in known breach datasets without revealing the passwords to the server. This application is a prime candidate for parallelisation as one could imagine the database being split into individual sections that are processed in parallel.

In the $\textbf{PIR-100}$ benchmark we evaluate the performance of our compiler for the PIR task on a database of size 100 where each entry is a 32-bit integer. In the $\textbf{PIR-500}$ benchmark we evaluate the performance of our compiler for the PIR task on a database of size 500 where each entry is a 32-bit integer. In \cref{tab:PIR100} and \cref{tab:PIR500} respectively we report the evaluation of the benchmark with different levels of parallelisation.

\begin{table}[ht]
\centering
\begin{tabular}{l|c|c}

& \textbf{Threads} & \textbf{Time} \\
\midrule
\servantoine   & 1   & 74s  \\
 \servantoine & 20     & 4.913s\\
 \servantoine & 50  & 2.26s \\
\servantoine & 100 &1.69s \\
\end{tabular}
\caption{The \textbf{PIR-100} benchmark.}
\label{tab:PIR100}
\end{table}

\begin{table}[ht]
\centering
\begin{tabular}{l|c|c}

& \textbf{Threads} & \textbf{Time} \\
\midrule
\servantoine   & 1   & 375.625s  \\
 \servantoine & 20     & 24.073s\\
 \servantoine & 50  & 10.360s \\
\servantoine & 100 &7.561s \\
\end{tabular}
\caption{The \textbf{PIR-500} benchmark.}
\label{tab:PIR500}
\end{table}
\section{Discussion}
We have implemented a compiler from Haskell to a simple boolean gate-based TFHE evaluation. 
As a next step the library will be extended to connect to more advanced FHE implementations such as tfhe-rs~\cite{tfhers} or FHE-Deck~\cite{fhedeck} which implement functional boostrapping which allows the evaluation of LUT tables.
We expect that the depth-based parallelization algorithm can achieve similar speedups for LUT-based circuits. 

%The possibility of parallelizing individual FHE operations is explored inthe OpenFHE and DEsilo libraries

Other approaches to FHE parallelization include parallelization of individual FHE operations targeting both CPU and GPU optimizations. This approach is implemented by the libraries OpenFHE~\cite{OpenFHE} and Desilo~\cite{Liberate_FHE}. It is an interesting question for future work to compare different parallelization techniques.

%But what do they actually do? Parallelising individual operations such as the bootstrapping? target CPU / GPU in particular?

\begin{acks}
We would like to thank Felix Klein from QBayLogic for helpful discussions regarding Clash.
\end{acks}

\bibliographystyle{ACM-Reference-Format}
\bibliography{references}


\begin{thebibliography}{00}

%%% the Web address.

\ifx \showCODEN    \undefined \def \showCODEN     #1{\unskip}     \fi
\ifx \showDOI      \undefined \def \showDOI       #1{#1}\fi
\ifx \showISBNx    \undefined \def \showISBNx     #1{\unskip}     \fi
\ifx \showISBNxiii \undefined \def \showISBNxiii  #1{\unskip}     \fi
\ifx \showISSN     \undefined \def \showISSN      #1{\unskip}     \fi
\ifx \showLCCN     \undefined \def \showLCCN      #1{\unskip}     \fi
\ifx \shownote     \undefined \def \shownote      #1{#1}          \fi
\ifx \showarticletitle \undefined \def \showarticletitle #1{#1}   \fi
\ifx \showURL      \undefined \def \showURL       {\relax}        \fi
% invisible to TeX
\providecommand\bibfield[2]{#2}
\providecommand\bibinfo[2]{#2}
\providecommand\natexlab[1]{#1}
\providecommand\showeprint[2][]{arXiv:#2}

\bibitem[\protect\citeauthoryear{Andreev and R{\"a}cke}{Andreev and R{\"a}cke}{2004}]%
        {AndreevHarald04}
\bibfield{author}{\bibinfo{person}{Konstantin Andreev} {and} \bibinfo{person}{Harald R{\"a}cke}.} \bibinfo{year}{2004}\natexlab{}.
\newblock \showarticletitle{Balanced graph partitioning}. In \bibinfo{booktitle}{{\em Proceedings of the sixteenth annual ACM symposium on Parallelism in algorithms and architectures}}. \bibinfo{pages}{120--124}.
\newblock


\bibitem[\protect\citeauthoryear{Archer, Calder{\'o}n~Trilla, Dagit, Malozemoff, Polyakov, Rohloff, and Ryan}{Archer et~al\mbox{.}}{2019}]%
        {ramparts}
\bibfield{author}{\bibinfo{person}{David~W Archer}, \bibinfo{person}{Jos{\'e}~Manuel Calder{\'o}n~Trilla}, \bibinfo{person}{Jason Dagit}, \bibinfo{person}{Alex Malozemoff}, \bibinfo{person}{Yuriy Polyakov}, \bibinfo{person}{Kurt Rohloff}, {and} \bibinfo{person}{Gerard Ryan}.} \bibinfo{year}{2019}\natexlab{}.
\newblock \showarticletitle{Ramparts: A programmer-friendly system for building homomorphic encryption applications}. In \bibinfo{booktitle}{{\em Proceedings of the 7th acm workshop on encrypted computing \& applied homomorphic cryptography}}. \bibinfo{pages}{57--68}.
\newblock


\bibitem[\protect\citeauthoryear{Badawi, Alexandru, Bates, Bergamaschi, Cousins, Erabelli, Genise, Halevi, Hunt, Kim, Lee, Liu, Micciancio, Pascoe, Polyakov, Quah, R.V., Rohloff, Saylor, Suponitsky, Triplett, Vaikuntanathan, and Zucca}{Badawi et~al\mbox{.}}{2022}]%
        {OpenFHE}
\bibfield{author}{\bibinfo{person}{Ahmad~Al Badawi}, \bibinfo{person}{Andreea Alexandru}, \bibinfo{person}{Jack Bates}, \bibinfo{person}{Flavio Bergamaschi}, \bibinfo{person}{David~Bruce Cousins}, \bibinfo{person}{Saroja Erabelli}, \bibinfo{person}{Nicholas Genise}, \bibinfo{person}{Shai Halevi}, \bibinfo{person}{Hamish Hunt}, \bibinfo{person}{Andrey Kim}, \bibinfo{person}{Yongwoo Lee}, \bibinfo{person}{Zeyu Liu}, \bibinfo{person}{Daniele Micciancio}, \bibinfo{person}{Carlo Pascoe}, \bibinfo{person}{Yuriy Polyakov}, \bibinfo{person}{Ian Quah}, \bibinfo{person}{Saraswathy R.V.}, \bibinfo{person}{Kurt Rohloff}, \bibinfo{person}{Jonathan Saylor}, \bibinfo{person}{Dmitriy Suponitsky}, \bibinfo{person}{Matthew Triplett}, \bibinfo{person}{Vinod Vaikuntanathan}, {and} \bibinfo{person}{Vincent Zucca}.} \bibinfo{year}{2022}\natexlab{}.
\newblock \bibinfo{title}{{OpenFHE}: Open-Source Fully Homomorphic Encryption Library}.
\newblock \bibinfo{howpublished}{Cryptology ePrint Archive, Paper 2022/915}.   (\bibinfo{year}{2022}).
\newblock
\showURL{%
\url{https://eprint.iacr.org/2022/915}}
\newblock
\shownote{\url{https://eprint.iacr.org/2022/915}.}


\bibitem[\protect\citeauthoryear{Bela{\"\i}d, Bon, Boudguiga, Sirdey, Trama, and Ye}{Bela{\"\i}d et~al\mbox{.}}{2025}]%
        {SNARDN25}
\bibfield{author}{\bibinfo{person}{Sonia Bela{\"\i}d}, \bibinfo{person}{Nicolas Bon}, \bibinfo{person}{Aymen Boudguiga}, \bibinfo{person}{Renaud Sirdey}, \bibinfo{person}{Daphn{\'e} Trama}, {and} \bibinfo{person}{Nicolas Ye}.} \bibinfo{year}{2025}\natexlab{}.
\newblock \showarticletitle{Further Improvements in AES Execution over TFHE: Towards Breaking the 1 sec Barrier}.
\newblock \bibinfo{journal}{{\em Cryptology ePrint Archive\/}} (\bibinfo{year}{2025}).
\newblock


\bibitem[\protect\citeauthoryear{Boemer, Lao, Cammarota, and Wierzynski}{Boemer et~al\mbox{.}}{2019}]%
        {nGraph-HE}
\bibfield{author}{\bibinfo{person}{Fabian Boemer}, \bibinfo{person}{Yixing Lao}, \bibinfo{person}{Rosario Cammarota}, {and} \bibinfo{person}{Casimir Wierzynski}.} \bibinfo{year}{2019}\natexlab{}.
\newblock \showarticletitle{nGraph-HE: a graph compiler for deep learning on homomorphically encrypted data}. In \bibinfo{booktitle}{{\em Proceedings of the 16th ACM international conference on computing frontiers}}. \bibinfo{pages}{3--13}.
\newblock


\bibitem[\protect\citeauthoryear{Bon, Pointcheval, and Rivain}{Bon et~al\mbox{.}}{2024}]%
        {BPR24}
\bibfield{author}{\bibinfo{person}{Nicolas Bon}, \bibinfo{person}{David Pointcheval}, {and} \bibinfo{person}{Matthieu Rivain}.} \bibinfo{year}{2024}\natexlab{}.
\newblock \showarticletitle{Optimized homomorphic evaluation of boolean functions}.
\newblock \bibinfo{journal}{{\em IACR Transactions on Cryptographic Hardware and Embedded Systems\/}} \bibinfo{volume}{2024}, \bibinfo{number}{3} (\bibinfo{year}{2024}), \bibinfo{pages}{302--341}.
\newblock


\bibitem[\protect\citeauthoryear{Boyar and Peralta}{Boyar and Peralta}{2012}]%
        {bpsbox}
\bibfield{author}{\bibinfo{person}{Joan Boyar} {and} \bibinfo{person}{Ren{\'e} Peralta}.} \bibinfo{year}{2012}\natexlab{}.
\newblock \showarticletitle{A Small Depth-16 Circuit for the AES S-Box}. In \bibinfo{booktitle}{{\em Information Security and Privacy Research}}, \bibfield{editor}{\bibinfo{person}{Dimitris Gritzalis}, \bibinfo{person}{Steven Furnell}, {and} \bibinfo{person}{Marianthi Theoharidou}} (Eds.). \bibinfo{publisher}{Springer Berlin Heidelberg}, \bibinfo{address}{Berlin, Heidelberg}, \bibinfo{pages}{287--298}.
\newblock


\bibitem[\protect\citeauthoryear{Brakerski}{Brakerski}{2012}]%
        {brakerski12}
\bibfield{author}{\bibinfo{person}{Zvika Brakerski}.} \bibinfo{year}{2012}\natexlab{}.
\newblock \showarticletitle{Fully homomorphic encryption without modulus switching from classical GapSVP}. In \bibinfo{booktitle}{{\em Annual cryptology conference}}. Springer, \bibinfo{pages}{868--886}.
\newblock


\bibitem[\protect\citeauthoryear{Brakerski, Gentry, and Vaikuntanathan}{Brakerski et~al\mbox{.}}{2014}]%
        {BGV}
\bibfield{author}{\bibinfo{person}{Zvika Brakerski}, \bibinfo{person}{Craig Gentry}, {and} \bibinfo{person}{Vinod Vaikuntanathan}.} \bibinfo{year}{2014}\natexlab{}.
\newblock \showarticletitle{(Leveled) fully homomorphic encryption without bootstrapping}.
\newblock \bibinfo{journal}{{\em ACM Transactions on Computation Theory (TOCT)\/}} \bibinfo{volume}{6}, \bibinfo{number}{3} (\bibinfo{year}{2014}), \bibinfo{pages}{1--36}.
\newblock


\bibitem[\protect\citeauthoryear{Carpov, Dubrulle, and Sirdey}{Carpov et~al\mbox{.}}{2015}]%
        {cingulata}
\bibfield{author}{\bibinfo{person}{Sergiu Carpov}, \bibinfo{person}{Paul Dubrulle}, {and} \bibinfo{person}{Renaud Sirdey}.} \bibinfo{year}{2015}\natexlab{}.
\newblock \showarticletitle{Armadillo: a compilation chain for privacy preserving applications}. In \bibinfo{booktitle}{{\em Proceedings of the 3rd International Workshop on Security in Cloud Computing}}. \bibinfo{pages}{13--19}.
\newblock


\bibitem[\protect\citeauthoryear{Cheon, Kim, Kim, and Song}{Cheon et~al\mbox{.}}{2017}]%
        {AC:CKKS17}
\bibfield{author}{\bibinfo{person}{Jung~Hee Cheon}, \bibinfo{person}{Andrey Kim}, \bibinfo{person}{Miran Kim}, {and} \bibinfo{person}{Yongsoo Song}.} \bibinfo{year}{2017}\natexlab{}.
\newblock \showarticletitle{Homomorphic Encryption for Arithmetic of Approximate Numbers}. In \bibinfo{booktitle}{{\em Advances in Cryptology -- ASIACRYPT 2017}}, \bibfield{editor}{\bibinfo{person}{Tsuyoshi Takagi} {and} \bibinfo{person}{Thomas Peyrin}} (Eds.). \bibinfo{publisher}{Springer International Publishing}, \bibinfo{address}{Cham}, \bibinfo{pages}{409--437}.
\newblock
\showISBNx{978-3-319-70694-8}


\bibitem[\protect\citeauthoryear{Chielle, Mazonka, Gamil, and Maniatakos}{Chielle et~al\mbox{.}}{2022a}]%
        {alchemy}
\bibfield{author}{\bibinfo{person}{Eduardo Chielle}, \bibinfo{person}{Oleg Mazonka}, \bibinfo{person}{Homer Gamil}, {and} \bibinfo{person}{Michail Maniatakos}.} \bibinfo{year}{2022}\natexlab{a}.
\newblock \showarticletitle{Accelerating fully homomorphic encryption by bridging modular and bit-level arithmetic}. In \bibinfo{booktitle}{{\em Proceedings of the 41st IEEE/ACM International Conference on Computer-Aided Design}}. \bibinfo{pages}{1--9}.
\newblock


\bibitem[\protect\citeauthoryear{Chielle, Mazonka, Gamil, and Maniatakos}{Chielle et~al\mbox{.}}{2022b}]%
        {E3}
\bibfield{author}{\bibinfo{person}{Eduardo Chielle}, \bibinfo{person}{Oleg Mazonka}, \bibinfo{person}{Homer Gamil}, {and} \bibinfo{person}{Michail Maniatakos}.} \bibinfo{year}{2022}\natexlab{b}.
\newblock \showarticletitle{Accelerating fully homomorphic encryption by bridging modular and bit-level arithmetic}. In \bibinfo{booktitle}{{\em Proceedings of the 41st IEEE/ACM International Conference on Computer-Aided Design}}. \bibinfo{pages}{1--9}.
\newblock


\bibitem[\protect\citeauthoryear{Chillotti, Gama, Georgieva, and Izabach{\`e}ne}{Chillotti et~al\mbox{.}}{2019}]%
        {TFHE}
\bibfield{author}{\bibinfo{person}{Ilaria Chillotti}, \bibinfo{person}{Nicolas Gama}, \bibinfo{person}{Mariya Georgieva}, {and} \bibinfo{person}{Malika Izabach{\`e}ne}.} \bibinfo{year}{2019}\natexlab{}.
\newblock \showarticletitle{TFHE: Fast Fully Homomorphic Encryption Over the Torus}.
\newblock \bibinfo{journal}{{\em Journal of Cryptology\/}}  \bibinfo{volume}{33} (\bibinfo{year}{2019}), \bibinfo{pages}{34 -- 91}.
\newblock
\showURL{%
\url{https://api.semanticscholar.org/CorpusID:44099955}}


\bibitem[\protect\citeauthoryear{Chillotti, Gama, Georgieva, and Izabach{\`e}ne}{Chillotti et~al\mbox{.}}{2016}]%
        {TFHElib}
\bibfield{author}{\bibinfo{person}{Ilaria Chillotti}, \bibinfo{person}{Nicolas Gama}, \bibinfo{person}{Mariya Georgieva}, {and} \bibinfo{person}{Malika Izabach{\`e}ne}.} \bibinfo{year}{August 2016}\natexlab{}.
\newblock \bibinfo{title}{{TFHE}: Fast Fully Homomorphic Encryption Library}.
\newblock   (\bibinfo{year}{August 2016}).
\newblock
\newblock
\shownote{https://tfhe.github.io/tfhe/.}


\bibitem[\protect\citeauthoryear{Cowan, Dangwal, Alaghi, Trippel, Lee, and Reagen}{Cowan et~al\mbox{.}}{2021}]%
        {porcupine}
\bibfield{author}{\bibinfo{person}{Meghan Cowan}, \bibinfo{person}{Deeksha Dangwal}, \bibinfo{person}{Armin Alaghi}, \bibinfo{person}{Caroline Trippel}, \bibinfo{person}{Vincent~T Lee}, {and} \bibinfo{person}{Brandon Reagen}.} \bibinfo{year}{2021}\natexlab{}.
\newblock \showarticletitle{Porcupine: A synthesizing compiler for vectorized homomorphic encryption}. In \bibinfo{booktitle}{{\em Proceedings of the 42nd ACM SIGPLAN International Conference on Programming Language Design and Implementation}}. \bibinfo{pages}{375--389}.
\newblock


\bibitem[\protect\citeauthoryear{Dathathri, Kostova, Saarikivi, Dai, Laine, and Musuvathi}{Dathathri et~al\mbox{.}}{2020}]%
        {eva}
\bibfield{author}{\bibinfo{person}{Roshan Dathathri}, \bibinfo{person}{Blagovesta Kostova}, \bibinfo{person}{Olli Saarikivi}, \bibinfo{person}{Wei Dai}, \bibinfo{person}{Kim Laine}, {and} \bibinfo{person}{Madan Musuvathi}.} \bibinfo{year}{2020}\natexlab{}.
\newblock \showarticletitle{EVA: An encrypted vector arithmetic language and compiler for efficient homomorphic computation}. In \bibinfo{booktitle}{{\em Proceedings of the 41st ACM SIGPLAN conference on programming language design and implementation}}. \bibinfo{pages}{546--561}.
\newblock


\bibitem[\protect\citeauthoryear{DESILO}{DESILO}{2023}]%
        {Liberate_FHE}
\bibfield{author}{\bibinfo{person}{DESILO}.} \bibinfo{year}{2023}\natexlab{}.
\newblock \bibinfo{title}{{Liberate.FHE: A New FHE Library for Bridging the Gap between Theory and Practice with a Focus on Performance and Accuracy}}.
\newblock   (\bibinfo{year}{2023}).
\newblock
\newblock
\shownote{\url{https://github.com/Desilo/liberate-fhe}.}


\bibitem[\protect\citeauthoryear{Ducas and Micciancio}{Ducas and Micciancio}{2015}]%
        {EC:DucMic15}
\bibfield{author}{\bibinfo{person}{L{\'e}o Ducas} {and} \bibinfo{person}{Daniele Micciancio}.} \bibinfo{year}{2015}\natexlab{}.
\newblock \showarticletitle{FHEW: Bootstrapping Homomorphic Encryption in Less Than a Second}. In \bibinfo{booktitle}{{\em Advances in Cryptology -- EUROCRYPT 2015}}, \bibfield{editor}{\bibinfo{person}{Elisabeth Oswald} {and} \bibinfo{person}{Marc Fischlin}} (Eds.). \bibinfo{publisher}{Springer Berlin Heidelberg}, \bibinfo{address}{Berlin, Heidelberg}, \bibinfo{pages}{617--640}.
\newblock
\showISBNx{978-3-662-46800-5}


\bibitem[\protect\citeauthoryear{Fan and Vercauteren}{Fan and Vercauteren}{2012}]%
        {FV}
\bibfield{author}{\bibinfo{person}{Junfeng Fan} {and} \bibinfo{person}{Frederik Vercauteren}.} \bibinfo{year}{2012}\natexlab{}.
\newblock \showarticletitle{Somewhat practical fully homomorphic encryption}.
\newblock \bibinfo{journal}{{\em Cryptology ePrint Archive\/}} (\bibinfo{year}{2012}).
\newblock


\bibitem[\protect\citeauthoryear{Gentry}{Gentry}{2009}]%
        {gentry09}
\bibfield{author}{\bibinfo{person}{Craig Gentry}.} \bibinfo{year}{2009}\natexlab{}.
\newblock \showarticletitle{Fully homomorphic encryption using ideal lattices}. In \bibinfo{booktitle}{{\em Proceedings of the forty-first annual ACM symposium on Theory of computing}}. \bibinfo{pages}{169--178}.
\newblock


\bibitem[\protect\citeauthoryear{Gentry, Halevi, and Smart}{Gentry et~al\mbox{.}}{2012}]%
        {C:CSN12}
\bibfield{author}{\bibinfo{person}{Craig Gentry}, \bibinfo{person}{Shai Halevi}, {and} \bibinfo{person}{Nigel~P. Smart}.} \bibinfo{year}{2012}\natexlab{}.
\newblock \showarticletitle{Homomorphic Evaluation of the AES Circuit}. In \bibinfo{booktitle}{{\em Advances in Cryptology -- CRYPTO 2012}}, \bibfield{editor}{\bibinfo{person}{Reihaneh Safavi-Naini} {and} \bibinfo{person}{Ran Canetti}} (Eds.). \bibinfo{publisher}{Springer Berlin Heidelberg}, \bibinfo{address}{Berlin, Heidelberg}, \bibinfo{pages}{850--867}.
\newblock


\bibitem[\protect\citeauthoryear{Gentry, Sahai, and Waters}{Gentry et~al\mbox{.}}{2013}]%
        {C:GSW13}
\bibfield{author}{\bibinfo{person}{Craig Gentry}, \bibinfo{person}{Amit Sahai}, {and} \bibinfo{person}{Brent Waters}.} \bibinfo{year}{2013}\natexlab{}.
\newblock \showarticletitle{Homomorphic Encryption from Learning with Errors: Conceptually-Simpler, Asymptotically-Faster, Attribute-Based}.
\newblock \bibinfo{journal}{{\em IACR Cryptol. ePrint Arch.\/}}  \bibinfo{volume}{2013} (\bibinfo{year}{2013}), \bibinfo{pages}{340}.
\newblock
\showURL{%
\url{https://api.semanticscholar.org/CorpusID:15191005}}


\bibitem[\protect\citeauthoryear{Gorantala, Springer, Purser-Haskell, Lam, Wilson, Ali, Astor, Zukerman, Ruth, Dibak, Schoppmann, Kulankhina, Forget, Marn, Tew, Misoczki, Guillen, Ye, Kraft, Desfontaines, Krishnamurthy, Guevara, Perera, Sushko, and Gipson}{Gorantala et~al\mbox{.}}{2021}]%
        {HEIR}
\bibfield{author}{\bibinfo{person}{Shruthi Gorantala}, \bibinfo{person}{Rob Springer}, \bibinfo{person}{Sean Purser-Haskell}, \bibinfo{person}{William Lam}, \bibinfo{person}{Royce Wilson}, \bibinfo{person}{Asra Ali}, \bibinfo{person}{Eric~P. Astor}, \bibinfo{person}{Itai Zukerman}, \bibinfo{person}{Sam Ruth}, \bibinfo{person}{Christoph Dibak}, \bibinfo{person}{Phillipp Schoppmann}, \bibinfo{person}{Sasha Kulankhina}, \bibinfo{person}{Alain Forget}, \bibinfo{person}{David Marn}, \bibinfo{person}{Cameron Tew}, \bibinfo{person}{Rafael Misoczki}, \bibinfo{person}{Bernat Guillen}, \bibinfo{person}{Xinyu Ye}, \bibinfo{person}{Dennis Kraft}, \bibinfo{person}{Damien Desfontaines}, \bibinfo{person}{Aishe Krishnamurthy}, \bibinfo{person}{Miguel Guevara}, \bibinfo{person}{Irippuge~Milinda Perera}, \bibinfo{person}{Yurii Sushko}, {and} \bibinfo{person}{Bryant Gipson}.} \bibinfo{year}{2021}\natexlab{}.
\newblock \bibinfo{title}{A General Purpose Transpiler for Fully Homomorphic Encryption}.
\newblock \bibinfo{howpublished}{Cryptology {ePrint} Archive, Paper 2021/811}.   (\bibinfo{year}{2021}).
\newblock
\showURL{%
\url{https://eprint.iacr.org/2021/811}}


\bibitem[\protect\citeauthoryear{Kannepalli, Laine, and Moreno}{Kannepalli et~al\mbox{.}}{2021}]%
        {psswordmonitor}
\bibfield{author}{\bibinfo{person}{Sreekanth Kannepalli}, \bibinfo{person}{Kim Laine}, {and} \bibinfo{person}{Radames~Cruz Moreno}.} \bibinfo{year}{2021}\natexlab{}.
\newblock \showarticletitle{Password monitor: Safeguarding passwords in microsoft edge}.
\newblock \bibinfo{journal}{{\em Microsoft Research Blog\/}} (\bibinfo{year}{2021}).
\newblock
\showURL{%
\url{https://www.microsoft.com/en-us/research/blog/password-monitor-safeguarding-passwords-in-microsoft-edge/}}


\bibitem[\protect\citeauthoryear{Kluczniak and Schild}{Kluczniak and Schild}{2024}]%
        {fhedeck}
\bibfield{author}{\bibinfo{person}{Kamil Kluczniak} {and} \bibinfo{person}{Leonard Schild}.} \bibinfo{year}{2024}\natexlab{}.
\newblock \bibinfo{title}{{FDFB}$^2$: Functional Bootstrapping via Sparse Polynomial Multiplication}.
\newblock \bibinfo{howpublished}{Cryptology {ePrint} Archive, Paper 2024/1376}.   (\bibinfo{year}{2024}).
\newblock
\showURL{%
\url{https://eprint.iacr.org/2024/1376}}


\bibitem[\protect\citeauthoryear{Krastev, Samardzic, Langowski, Devadas, and Sanchez}{Krastev et~al\mbox{.}}{2024}]%
        {fhelipe}
\bibfield{author}{\bibinfo{person}{Aleksandar Krastev}, \bibinfo{person}{Nikola Samardzic}, \bibinfo{person}{Simon Langowski}, \bibinfo{person}{Srinivas Devadas}, {and} \bibinfo{person}{Daniel Sanchez}.} \bibinfo{year}{2024}\natexlab{}.
\newblock \showarticletitle{A Tensor Compiler with Automatic Data Packing for Simple and Efficient Fully Homomorphic Encryption}.
\newblock \bibinfo{journal}{{\em Proc. ACM Program. Lang.\/}} \bibinfo{volume}{8}, \bibinfo{number}{PLDI}, Article \bibinfo{articleno}{152} (\bibinfo{date}{June} \bibinfo{year}{2024}), \bibinfo{numpages}{25}~pages.
\newblock
\showDOI{%
\url{https://doi.org/10.1145/3656382}}


\bibitem[\protect\citeauthoryear{Li, Lai, Yuan, Sui, Liu, Zhu, Zhang, Xiao, Chen, and Xue}{Li et~al\mbox{.}}{2025}]%
        {antace}
\bibfield{author}{\bibinfo{person}{Long Li}, \bibinfo{person}{Jianxin Lai}, \bibinfo{person}{Peng Yuan}, \bibinfo{person}{Tianxiang Sui}, \bibinfo{person}{Yan Liu}, \bibinfo{person}{Qing Zhu}, \bibinfo{person}{Xiaojing Zhang}, \bibinfo{person}{Linjie Xiao}, \bibinfo{person}{Wenguang Chen}, {and} \bibinfo{person}{Jingling Xue}.} \bibinfo{year}{2025}\natexlab{}.
\newblock \showarticletitle{ANT-ACE: An FHE Compiler Framework for Automating Neural Network Inference}. In \bibinfo{booktitle}{{\em Proceedings of the 23rd ACM/IEEE International Symposium on Code Generation and Optimization}}. \bibinfo{pages}{193--208}.
\newblock


\bibitem[\protect\citeauthoryear{Mella and Susella}{Mella and Susella}{2013}]%
        {MS13}
\bibfield{author}{\bibinfo{person}{Silvia Mella} {and} \bibinfo{person}{Ruggero Susella}.} \bibinfo{year}{2013}\natexlab{}.
\newblock \showarticletitle{On the Homomorphic Computation of Symmetric Cryptographic Primitives}. In \bibinfo{booktitle}{{\em Cryptography and Coding}}, \bibfield{editor}{\bibinfo{person}{Martijn Stam}} (Ed.). \bibinfo{publisher}{Springer Berlin Heidelberg}, \bibinfo{address}{Berlin, Heidelberg}, \bibinfo{pages}{28--44}.
\newblock


\bibitem[\protect\citeauthoryear{Mono, Kluczniak, and G{\"u}neysu}{Mono et~al\mbox{.}}{2024}]%
        {MKG24}
\bibfield{author}{\bibinfo{person}{Johannes Mono}, \bibinfo{person}{Kamil Kluczniak}, {and} \bibinfo{person}{Tim G{\"u}neysu}.} \bibinfo{year}{2024}\natexlab{}.
\newblock \showarticletitle{Improved Circuit Synthesis with Multi-Value Bootstrapping for FHEW-like Schemes}.
\newblock \bibinfo{journal}{{\em IACR Transactions on Cryptographic Hardware and Embedded Systems\/}} \bibinfo{volume}{2024}, \bibinfo{number}{4} (\bibinfo{year}{2024}), \bibinfo{pages}{633--656}.
\newblock


\bibitem[\protect\citeauthoryear{Stracovsky, Mahdavi, and Kerschbaum}{Stracovsky et~al\mbox{.}}{2022}]%
        {stracovsky22}
\bibfield{author}{\bibinfo{person}{Roy Stracovsky}, \bibinfo{person}{Rasoul~Akhavan Mahdavi}, {and} \bibinfo{person}{Florian Kerschbaum}.} \bibinfo{year}{2022}\natexlab{}.
\newblock \showarticletitle{Faster evaluation of AES using TFHE}.
\newblock \bibinfo{journal}{{\em Poster Session, FHE. Org\/}} (\bibinfo{year}{2022}).
\newblock


\bibitem[\protect\citeauthoryear{Streit, Hammacher, Zeller, and Hack}{Streit et~al\mbox{.}}{2012}]%
        {SHZH12}
\bibfield{author}{\bibinfo{person}{Kevin Streit}, \bibinfo{person}{Clemens Hammacher}, \bibinfo{person}{Andreas Zeller}, {and} \bibinfo{person}{Sebastian Hack}.} \bibinfo{year}{2012}\natexlab{}.
\newblock \showarticletitle{Sambamba: A runtime system for online adaptive parallelization}. In \bibinfo{booktitle}{{\em Compiler Construction: 21st International Conference, CC 2012, Held as Part of the European Joint Conferences on Theory and Practice of Software, ETAPS 2012, Tallinn, Estonia, March 24--April 1, 2012. Proceedings 21}}. Springer, \bibinfo{pages}{240--243}.
\newblock


\bibitem[\protect\citeauthoryear{Trama, Clet, Boudguiga, and Sirdey}{Trama et~al\mbox{.}}{2023}]%
        {TCBS23}
\bibfield{author}{\bibinfo{person}{Daphn\'{e} Trama}, \bibinfo{person}{Pierre-Emmanuel Clet}, \bibinfo{person}{Aymen Boudguiga}, {and} \bibinfo{person}{Renaud Sirdey}.} \bibinfo{year}{2023}\natexlab{}.
\newblock \showarticletitle{A Homomorphic AES Evaluation in Less than 30 Seconds by Means of TFHE}. In \bibinfo{booktitle}{{\em Proceedings of the 11th Workshop on Encrypted Computing \& Applied Homomorphic Cryptography}} {\em (\bibinfo{series}{WAHC '23})}. \bibinfo{publisher}{Association for Computing Machinery}, \bibinfo{address}{New York, NY, USA}, \bibinfo{pages}{79–90}.
\newblock
\showISBNx{9798400702556}
\showDOI{%
\url{https://doi.org/10.1145/3605759.3625260}}


\bibitem[\protect\citeauthoryear{Viand, Jattke, Haller, and Hithnawi}{Viand et~al\mbox{.}}{2022}]%
        {heco}
\bibfield{author}{\bibinfo{person}{Alexander Viand}, \bibinfo{person}{Patrick Jattke}, \bibinfo{person}{Miro Haller}, {and} \bibinfo{person}{Anwar Hithnawi}.} \bibinfo{year}{2022}\natexlab{}.
\newblock \showarticletitle{HECO: Automatic code optimizations for efficient fully homomorphic encryption}.
\newblock \bibinfo{journal}{{\em arXiv preprint arXiv:2202.01649\/}} (\bibinfo{year}{2022}).
\newblock


\bibitem[\protect\citeauthoryear{Viand, Jattke, and Hithnawi}{Viand et~al\mbox{.}}{2021}]%
        {SOK}
\bibfield{author}{\bibinfo{person}{Alexander Viand}, \bibinfo{person}{Patrick Jattke}, {and} \bibinfo{person}{Anwar Hithnawi}.} \bibinfo{year}{2021}\natexlab{}.
\newblock \showarticletitle{SoK: Fully Homomorphic Encryption Compilers}. In \bibinfo{booktitle}{{\em 2021 IEEE Symposium on Security and Privacy (SP)}}. \bibinfo{pages}{1092--1108}.
\newblock
\showDOI{%
\url{https://doi.org/10.1109/SP40001.2021.00068}}


\bibitem[\protect\citeauthoryear{Viand and Shafagh}{Viand and Shafagh}{2018}]%
        {marble}
\bibfield{author}{\bibinfo{person}{Alexander Viand} {and} \bibinfo{person}{Hossein Shafagh}.} \bibinfo{year}{2018}\natexlab{}.
\newblock \showarticletitle{Marble: Making fully homomorphic encryption accessible to all}. In \bibinfo{booktitle}{{\em Proceedings of the 6th workshop on encrypted computing \& applied homomorphic cryptography}}. \bibinfo{pages}{49--60}.
\newblock


\bibitem[\protect\citeauthoryear{Vos, Conti, and Erkin}{Vos et~al\mbox{.}}{2024}]%
        {oraqle}
\bibfield{author}{\bibinfo{person}{Jelle Vos}, \bibinfo{person}{Mauro Conti}, {and} \bibinfo{person}{Zekeriya Erkin}.} \bibinfo{year}{2024}\natexlab{}.
\newblock \showarticletitle{Oraqle: A Depth-Aware Secure Computation Compiler}. In \bibinfo{booktitle}{{\em Proceedings of the 12th Workshop on Encrypted Computing \& Applied Homomorphic Cryptography}} {\em (\bibinfo{series}{WAHC '24})}. \bibinfo{publisher}{Association for Computing Machinery}, \bibinfo{address}{New York, NY, USA}, \bibinfo{pages}{43–50}.
\newblock
\showISBNx{9798400712418}
\showDOI{%
\url{https://doi.org/10.1145/3689945.3694808}}


\bibitem[\protect\citeauthoryear{Wei, Lu, Wang, Liu, Li, and Wang}{Wei et~al\mbox{.}}{2024}]%
        {thunderbird}
\bibfield{author}{\bibinfo{person}{Benqiang Wei}, \bibinfo{person}{Xianhui Lu}, \bibinfo{person}{Ruida Wang}, \bibinfo{person}{Kun Liu}, \bibinfo{person}{Zhihao Li}, {and} \bibinfo{person}{Kunpeng Wang}.} \bibinfo{year}{2024}\natexlab{}.
\newblock \showarticletitle{Thunderbird: Efficient Homomorphic Evaluation of Symmetric Ciphers in 3GPP by combining two modes of TFHE}.
\newblock \bibinfo{journal}{{\em IACR Transactions on Cryptographic Hardware and Embedded Systems\/}} \bibinfo{volume}{2024}, \bibinfo{number}{3} (\bibinfo{year}{2024}), \bibinfo{pages}{530--573}.
\newblock


\bibitem[\protect\citeauthoryear{Wei, Wang, Li, Liu, and Lu}{Wei et~al\mbox{.}}{2023}]%
        {fregata}
\bibfield{author}{\bibinfo{person}{Benqiang Wei}, \bibinfo{person}{Ruida Wang}, \bibinfo{person}{Zhihao Li}, \bibinfo{person}{Qinju Liu}, {and} \bibinfo{person}{Xianhui Lu}.} \bibinfo{year}{2023}\natexlab{}.
\newblock \showarticletitle{Fregata: Faster Homomorphic Evaluation of AES via TFHE}. In \bibinfo{booktitle}{{\em Information Security: 26th International Conference, ISC 2023, Groningen, The Netherlands, November 15–17, 2023, Proceedings}}. \bibinfo{publisher}{Springer-Verlag}, \bibinfo{address}{Berlin, Heidelberg}, \bibinfo{pages}{392–412}.
\newblock
\showISBNx{978-3-031-49186-3}
\showDOI{%
\url{https://doi.org/10.1007/978-3-031-49187-0_20}}


\bibitem[\protect\citeauthoryear{Zama}{Zama}{2022}]%
        {tfhers}
\bibfield{author}{\bibinfo{person}{Zama}.} \bibinfo{year}{2022}\natexlab{}.
\newblock \bibinfo{title}{{TFHE-rs: A Pure Rust Implementation of the TFHE Scheme for Boolean and Integer Arithmetics Over Encrypted Data}}.
\newblock   (\bibinfo{year}{2022}).
\newblock
\newblock
\shownote{\url{https://github.com/zama-ai/tfhe-rs}.}


\end{thebibliography}

\end{document}